# Strengthening Power System Resilience to Extreme Weather Events Through Grid Enhancing Technologies

Joseph Nyangon, Ph.D.*, Senior Member, IEEE, U.S. Department of Energy, Washington, DC., U.S.A.

*Abstract*— Climate change significantly increases risks to power systems, exacerbating issues such as aging infrastructure, evolving regulations, cybersecurity threats, and fluctuating demand. This paper focuses on the utilization of Grid Enhancing Technologies (GETs) to strengthen power system resilience in the face of extreme weather events. GETs are pivotal in optimizing energy distribution, enabling predictive maintenance, ensuring reliable electricity supply, facilitating renewable energy integration, and automating responses to power instabilities and outages. Drawing insights from resilience theory, the paper reviews recent grid resilience literature, highlighting increasing vulnerabilities due to severe weather events. It demonstrates how GETs are crucial in optimizing smart grid operations, thereby not only mitigating climate-related impacts but also promoting industrial transformation.

## I. Introduction

The escalating frequency and severity of extreme weather events, exacerbated by climate change, underscore the vulnerability of global energy systems. The 2021 devastating Uri winter storm and Hurricane Maria in 2017, have starkly exposed the susceptibility of electricity infrastructure to climate-induced challenges [1], [2]. The IPCC's Sixth Assessment Report (2023) highlighted these vulnerabilities, emphasizing the urgent need for enhanced resilience in electrical power grids [3].

Studies show that integrating grid-enhancing technologies (GETs) like dynamic line ratings (DLR), flexible alternating current transmission system (FACTS), and topology optimization (TO) can unlock additional capacity on the existing transmission and distribution (T&D) systems, and improve power system resilience [4]–[6]. DLR, for instance, real-time forecasting of current-carrying capacity of transmission lines, facilitating better grid planning and resource allocation under various conditions. FACTS, utilizing artificial intelligence and machine learning (AI/ML) powered algorithms and distributed energy resources (DERs) like distributed generation, energy storage, and demand response, offers localized support during grid disturbances. On the other hand, TO, through the use of microgrids, SCADA (Supervisory Control and Data Acquisition) systems, AMI (Advanced Metering Infrastructure), ADMS (Advanced Distribution Management Systems), integrates various grid management functions, enabling effective response to outages and disturbances. Additionally, deploying smart grids and long duration energy storage (LDES) can mitigate power disruptions during extended outages or emergencies, thus enhancing grid resilience.

GETs play a pivotal role in enhancing grid resilience, contributing to energy security, and mitigating climate change impacts on vulnerable communities [5]. They complement the utility of T&D infrastructure, enhancing the effectiveness and cost-efficiency of new T&D investments. The benefits of GETs are evident before, during, and after the development of traditional T&D projects [6]. However, it is important to distinguish between grid resilience and reliability. Reliability focuses on minor, localized disruptions, whereas resilience is about the system's ability to withstand or recover from larger, widespread events [7]. Traditional reliability metrics like System Average Interruption Duration Index (SAIDI), System Average Interruption Frequency Index (SAIFI) and Customer Average Interruption Duration Index (CAIDI) are essential for assessing electricity distribution system performance, but they may not fully capture the complexities of grid resilience [8]. Thus, an integrated approach that combines mitigation and adaptation strategies is essential to holistically address the impacts of climate change on energy infrastructure. This approach should account for the complexity and interconnected nature of electric power systems, considering the diverse end-use sectors, technologies, timeframes, and geographies [5].

## II. Conceptualizing Power System RESILIENCE

### A. Fundamentals of Resilience Theory in Power Systems

Resilience theory focuses on a system's ability to adapt and maintain functionality amidst disturbances, integrating insights from ecology, psychology, and complexity science [7], [9]. It advocates for an integrated approach in energy systems, emphasizing adaptability, redundancy, and learning to enhance resilience against climate change impacts [10]. Helmrich et al. [11] and Gilrein et al. [12] caution against conflating robustness with resilience, advocating for flexible, multifunctional infrastructure that accommodates evolving needs and the interconnectedness of system components. This holistic perspective is vital for managing complex power systems, where accurate risk assessment remains challenging [2].

### B. Resilience Trapezoid Curve and Curve

In the realm of energy infrastructure resilience, the resilience trapezoid and curve, as outlined by Mishra et al. [13] and Panteli et al. [14], depict an electrical grid's response to extreme weather through five stages: pre-disturbance normalcy (R0), resilience drop during disturbance ($R$pd), post-disturbance degradation, restoration, and final resilience level, which may not fully recover to R0. This framework's "resilience deficit" quantifies the impact of disruptions. Integrating resilience theory with grid hardening strategies, as suggested by Hosseini & Parvania [15] and Tsuchida et al. [6] requires technologies that ensure flexibility and dynamism like DLR, FACTS and TO. The modularity and scalability of these hardening strategies can be enhanced further by leveraging AI- and ML-techniques and large language models [16], [8].

2024. This research received partial support from the National Science Foundation (NSF) and the U.S. Department of Energy (DOE) under NSF CA No. EEC-1041895. Any opinions, findings and conclusions or recommendations expressed in this material are those of the author and do not necessarily reflect those of NSF or DOE. * Email contact: jnyangon@udel.edu

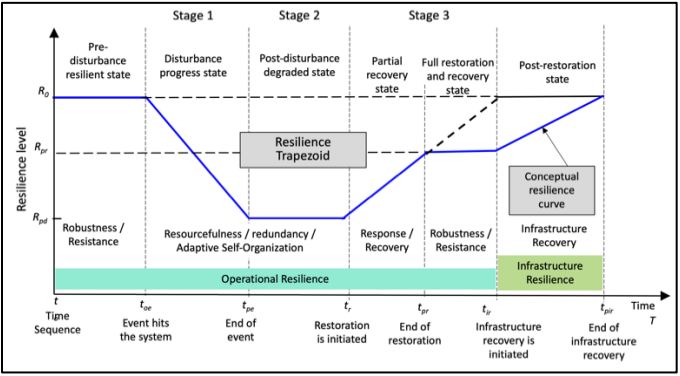

Fig. 1. Conceptual resilience trapezoid and curve for with an event, modified Panteli et al. [17].

In structural engineering, phenomena like infrastructure failure and material strength properties often follow a lognormal distribution. This distribution is used to calculate the likelihood of an electricity infrastructure reaching or surpassing a specific damage state (*ds*) under a given hazard intensity (e.g., impact on transformers or substations at certain damage thresholds). The probability that the damage state exceeds a certain threshold (*ds*) for a given hazard intensity can be expressed as:

$$P(X \geq ds) = 1 - P(X < ds)$$

where X is the random variable representing the damage state. Assuming X follows a lognormal distribution, the cumulative distribution function (CDF) of X is given by:

$$F_x(x) = \frac{1}{2} + \frac{1}{2}\phi\left[\frac{\ln(x) - \mu}{\sqrt[\sigma]{2}}\right]$$

where: $\mu$ is the mean and $\sigma$ is the standard deviation of the lognormal distribution at which the asset reaches the threshold of the damage state *ds*; *ds* is the damage state as a quantitative measure of damage (e.g., degree of cracking, deformation, etc.). The probability of an infrastructure being in or exceeding the damage state *ds* is then given by:

$$P(X \geq ds) = 1 - F_x(ds)$$

### III. Disruptive Impacts of Climate Change on Electricity Systems

Climate change presents challenges to electricity generation, T&D, and demand. Understanding these impacts and associated risks is needed to develop effective hardening strategies.

*A. Electricity Generation Infrastructure*

Climate change significantly impacts electricity generation infrastructure and assets. For example, hydroelectric power is vulnerable to altered river runoff due to changing temperatures and precipitation. Solar PV outputs decline with higher temperatures and increased cloud cover while wind turbines may suffer from infrastructure damages caused by intensified storms. Coastal energy infrastructures also face risks from climate-related disasters like floods and erosion. Thermoelectric plants, including fossil, biomass, and nuclear, struggle with cooling water scarcity and higher temperatures, leading to necessary adaptations like cooling system conversions [18]. In addition, extreme weather conditions threaten nuclear plant safety and disrupts raw material availability for thermal power.

*B. Electricity T&D Infrastructure*

Climate change also impacts T&D infrastructure, leading to increased line resistance, sagging, and service interruptions during heat waves [19]. Low wind speeds exacerbate these issues by reducing line cooling. Historical 2003 Northeast blackouts in the U.S. highlight the vulnerability of T&D systems to climate-induced risks like extreme weather, emphasizing the need for investment in climate-resilient infrastructure to proactively address potential transmission issues, alleviate congestion and integrate larger amounts of renewable resources [6]. The aging electricity infrastructure, facing climate-induced outages, incurs repair costs of $44 billion annually. Investment in GETs could cut these costs by over 50% [20].

*C. Electricity Demand and Consuption*

Climate change also reshapes electricity demand and consumption. Rising temperatures increase cooling demand, straining energy supply and risking outage. By 2050, cooling demand could rise by 30%, with a 2% increase in overall energy demand [21]. Strategies like energy-efficient cooling towers, demand-side management (DSM) programs and AI/ML-powered optimization can mitigate these challenges, promoting a resilient grid system against climate change impacts [22].

### IV. GET, DERs, LDES, DSM, and Microgrids

*A. GETs: DLR, FACTS and TO*

GETs, through DLR, FACTS, and TO, significantly contribute to the resilience and efficiency of power systems, especially in the face of extreme weather conditions induced by climate change. DLR adjusts the capacity of power lines in real-time, based on environmental conditions like temperature and wind speed, optimizing line usage and reducing congestion [6]. This approach contrasts with traditional static line ratings, which use conservative, fixed ampacity levels based on worst-case weather scenarios. A DLR system comprises sensors on transmission lines, a communication network, data analytics, and interfaces with energy management systems (EMSs) and SCADA for informed decision-making. Figure 2 illustrates a conceptual DLR system. FACTS, on the other hand, dynamically controls power flow and voltage stability, redistributing power to maximize line utilization without altering generator output [5]. On the other hand, TO employs AI-driven software to reroute power flows, alleviating bottlenecks and optimizing transmission capabilities by adjusting grid configurations [5]. These technologies collectively enhance grid efficiency, offering flexible responses to varying operational demands and environmental conditions.

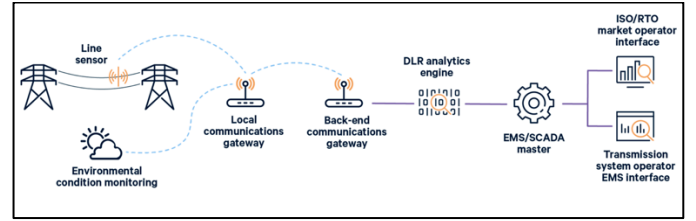

Fig. 2. Conceptual DLR system. Source: [5], 7.

*B. DERs*

DERs, including solar PV systems, wind turbines, and energy storage systems, are crucial for enhancing power grid resilience [23]. DERs improve energy efficiency, reliability, and flexibility, reducing reliance on centralized power sources [4], [24]. However, challenges like technology standardization and regulatory barriers persist. Effective DER management requires advanced metering and demand response programs. Texas' response to the 2021 grid failure highlights the importance of DERs in managing extreme events and adapting to climate change impacts [25]. DERs also contribute to climate change mitigation by enabling renewable energy integration and reducing greenhouse gas emissions. At both T&D levels, devises like phasor measurement units (PMUs) and high-speed transfer switches can help ensure grid stability during outages. Embracing DERs is key to sustainable energy development and climate resilience.

*C. LDES*

LDES is crucial for enhancing power systems' resilience by storing and discharging energy over extended periods, from hours to days. It is key in integrating renewable energy sources like wind and solar, which are intermittent and climate-dependent, into the grid. LDES provides system-wide benefits, including energy arbitrage, improved generator efficiency, and reduced costs, while managing transmission congestion and supporting grid resilience [26]. Diverse LDES methods like pumped water, compressed air, and batteries provide essential flexibility to manage the variable output of renewables. Additionally, integrating energy storage with smart grid technologies can harmonize electricity demand and supply, further enhancing grid resilience. Vehicle-to-grid technology also contributes to intelligent grid management and climate mitigation.

*D. DSM*

DSM is crucial for enhancing power systems resilience, reducing carbon intensity, and combating climate change in the electricity sector [27]. DSM strategies, including dynamic pricing and energy consumption feedback, promote energy conservation and awareness, thereby fostering climate resilience. Enhancements in transmission and distribution infrastructure, such as substation automation, also reinforce energy infrastructure resilience. Demand response, a key aspect of DSM, offers services like operating reserves and network capacity margin, extending beyond mere energy consumption. The integration of renewables and energy storage, coupled with initiatives like automated demand response, improves grid reliability and reduces greenhouse gas emissions. Power quality management and time-of-use programs maintain voltage stability and encourage off-peak energy use. Customer education and engagement in behavioral changes further strengthen grid resilience. Integrated operational resource management, combined with direct load control and real-time communication between energy management systems, optimizes performance and enhances network security.

*E. Microgrids*

Microgrids, self-sufficient energy systems, significantly boost power system resilience, particularly against climate-induced disruptions and power imbalances. These systems, capable of independent generation and storage, are crucial for high-reliability areas like hospitals and military bases [28]. They maintain functionality even during broader network failures through "islanding" operations, as evidenced by their performance in Tokyo during the 2011 earthquake and in the U.S. post-Hurricane Sandy [29]. Microgrids, integrating renewable sources and advanced technologies like AI, not only ensure continuous power during outages but also support sustainable communities and efficient energy trading [30]. The integration of microgrids at customer and community levels is essential for future power system resilience and efficiency.

*F. AI, ML, and SCADA Systems*

The integration of ADMS with SCADA systems, enhanced by AI and ML technologies, significantly improves power system resilience, particularly during extreme weather events. SCADA systems, equipped with AI algorithms, can proactively adjust power loads and predict flood-prone substations, enabling remote shutdowns or load redistributions to maintain power supply during severe weather. This integration is crucial for real-time data management, such as monitoring hydroelectric dam water levels and adjusting power generation accordingly. AI and ML technologies, including expert systems, fuzzy logic, and artificial neural networks, play a key role in designing, simulating, controlling, and ensuring fault tolerance in smart grids and renewable energy systems [31], [32].

Research highlights AI and ML's potential in enhancing grid resilience through decision-making in uncertain, high-dimensional spaces, damage detection, cyber-physical anomaly detection, and cybersecurity [15]. These GETs automate voltage regulation, optimize power system dynamics, and improve situational awareness and stability in power grids [33]. They also support the management of stochastic operations in T&D.

AI, ML, and SCADA technologies, when integrated, provide a robust framework for power systems to predict, respond, and recover from disruptions. Their ability to process vast datasets in real-time, coupled with SCADA's monitoring capabilities, can enhance resilience against extreme weather events.

V. INTEGRATING GETs WITH CLIMATE-RESILIENT STRATEGIES

*A. GETs and Smart Grids for Climate Resilience*

In the face of climate change, integrating smart grid technologies and GETs is vital for enhancing the resilience of power systems against extreme weather events. Smart grids, utilizing systems like AMI, can enhance grid resilience through real-time communication and swift outage responses. They also support diverse power generation and environmental sustainability, aligning with initiatives like the Biden Administration's renewable energy targets. This integration, encompassing grid hardening and advanced analytics, mitigates grid stress and emissions, leveraging renewable energy, energy storage, and microgrids [28], [34]. However, challenges in policy, market, and technology impede progress [25].

GETs are vital for maximizing existing grid capacity, offering cost-effective alternatives to new transmission projects [6]. Despite their scalability and portability, GETs face hurdles in market and regulatory structures, lacking incentives for widespread adoption [5]. For example, the current regulatory framework favors large capital investments, whereas GETs typically require smaller investments, with benefits accruing directly to consumers rather than owners. FERC order 2023 encourages investigation of alternative transmission technologies but stops short of mandating specific implementations [35]. A coordinated effort is necessary to evaluate and deploy the most effective smart grid solutions or GETs to enhance grid resilience and adapt to climate change.

*B. Leveraging AI and ML to Improve Grid Resilience*

Figure 2 illustrates deployment of various GETs in fault detection and management to improve situational awareness and operational remedies.

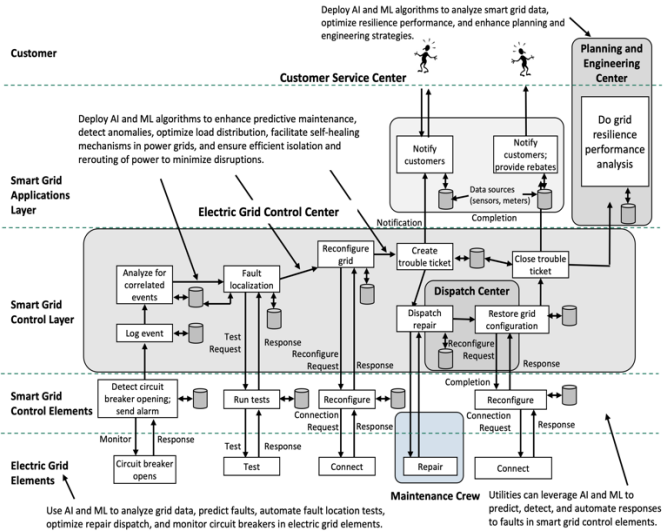

Fig. 3. An example showing deployment of GETs (AI, smart grids) in fault detection and management process to provide situational awareness and operational remedies

The integration of GETs ensures modular, resilient, and efficient power infrastructure (provided through situational awareness and operational remedies). Examples include:

- *Leverage AI and ML in Energy Forecasting and Optimization:* GETs can significantly boost the accuracy of energy forecasting and optimization. This leads to cost-effective grid management and enhanced energy distribution and storage. *Benefits*: GETs, like topology control software, can reduce outage impacts, manage energy demand, and increase renewable energy integration.

- *Use GETs in Weather-Related Grid Management*: Function: Smart grids, integrated with AI and ML, proactively predict and respond to extreme weather events like hurricanes. *Impact*: Studies by [13], [17], [32] emphasize GETs' role in efficient disaster response, anomaly detection, and grid behavior simulation under varying conditions, highlighting their importance in adapting to climate change.

Table 1 summarizes application of GETs in power system resilience under extreme weather events.

TABLE 1: SUMMARY OF COMPLEMENTATY BENEFITS OF GET SOLUTIONS THEIR TO GRID RESILIENCE

| GETs and Other Strategies | Potential Data Sources | Complementary Benefits of GETs to Grid Resilience |
|---|---|---|
| DLR | Weather data, Real-time line monitoring, Historical performance data | Increased transmission capacity and efficiency; Reduced congestion; Enhanced situational awareness |
| FACTS | System load data, Voltage and current measurements, Grid stability indicators | Improved power quality; Enhanced control over power flows; Increased system stability |
| TO powered by AI | Network configuration data; Load demand data; Grid performance metrics. | Optimized power flow paths; Reduced losses and operational costs; Enhanced flexibility in grid management. |
| Integration of ADMS with SCADA systems | SCADA, IoT sensors, weather forecasts, and power loads data. | Predict flood-prone substations, remote shutdowns, DLR, real-time grid monitoring, fault detection. |
| Smart grids with AMI for proactive response. | Real-time power consumption, fault detection, and historical hurricane data. | Predict vulnerabilities, reroute power, load reduction, targeted shutdowns. |
| Outage and energy forecasting (AI/ML); LDES optimization | Historical data; real-time energy demand and supply. | Accurate forecasting of electricity demand and supply; Predict and prevent equipment failures; maximize LDES capacity. |
| DSM, e.g., demand response | Smart meters, user preferences. | DLR, load shedding during peak demand or extreme events, improved grid stability. |
| PMU analysis with AI | PMU data. | Real-time grid monitoring, detecting disturbances, and improving grid stability. |

## VI. CONCLUSION

Electricity systems face regulatory pressure to minimize its environmental impact, rising energy demands, outdated grids, and climate change. Modernizing energy infrastructure, including the integration of GETs, is crucial for adapting to these challenges. GETs can enable the utility industry assess the disruptive impacts of climate change on electricity infrastructure and the economic value of the associated risks, including stranded asset risks due to the energy transition process.

However, GETs alone are insufficient. Strong legislative and regulatory frameworks are necessary to ensure grid modernization aligns with climate mitigation and adaptation goals. GETs, including DLR, FACTS and TO as well as smart grids, LDES, DLR, and DSM are vital for providing situational awareness and operational remedies during extreme situations triggered by climate change. GETs not only enable predictive maintenance, improve renewable energy forecasting, and cybersecurity, but also ensure modularity and scalability advantage ideal for ensuring power system resilience and addressing unintended congestion in T&D networks.

ACKNOWLEDGMENT

The author acknowledges valuable inputs from anonymous reviewers, which significantly improved this research.